\documentclass[superscriptaddress,amsmath, nofootinbib]{revtex4}
\usepackage{amsfonts}
\usepackage{graphicx}
\usepackage[brazil]{babel}
\usepackage[latin1]{inputenc}
\usepackage{amsmath}
\usepackage{amssymb}
\begin{document}


\title{Poincar\'e-Chetaev equations in the Dirac's formalism of constrained systems.}

\author{Alexei A. Deriglazov }
\email{alexei.deriglazov@ufjf.br} \affiliation{Depto. de Matem\'atica, ICE, Universidade Federal de Juiz de Fora,
MG, Brazil} 

\author{}

\date{\today}

\begin{abstract}
We single out a class of Lagrangians on a group manifold, for which one can introduce non-canonical coordinates in the phase space, which simplify the construction of the Poisson structure without explicitly calculating the Dirac bracket. In the case of $SO(3)$\,- manifold, the application of this formalism leads to the Poincar\'e-Chetaev equations. The general solution to these equations is written in terms of exponential of the Hamiltonian vector field. 
\end{abstract}

\maketitle 



\section{Introduction.}

The Euler-Poisson equations represent a very interesting example of a Hamiltonian system on the group manifold $SO(3)$, in which the number of conjugate momenta is less than the number of coordinates, and endowed with a non-canonical Poisson structure. Generalization of these  equations to other group manifolds are known as the Poincar\'e-Chetaev equations, and represent an active field of research \cite{Hol_2007, Hol_1999, Dub_2001, Boy_2005}. They were suggested by Poincar\'e in the Lagrangian-like form in \cite{Poi_1901}, while their Hamiltonian character was recognized by Chetaev in \cite{Chet_1941}. For the case of $SO(3)$, the Euler-Poisson equations can be obtained using the Dirac's formalism for constrained systems \cite{Dir_1950,GT,deriglazov2010classical}, by Hamiltonization the Lagrangian variational problem for a rigid body considered as a system with kinematic constraints \cite{AAD23, AAD23_1,AAD23_5}. Since this works adequately  
for $SO(3)$\,-manifold, it is interesting to see, whether it can be applied for other group manifolds and, more generally, for an arbitrary curved manifold.  This is the goal of the present work. We show how the Poincar\'e-Chetaev equations for certain class of Lagrangians on a curved manifold can be obtained following the Dirac's formalism.  

Let's start from a little more detailed description of the problem. 

Consider a mechanical system with nonsingular Lagrangian $L(q^A, \dot q^A)$, defined on configuration space with the coordinates $q^A(t)$, $A=1, 2, \ldots n$. Suppose the "particle" $q^A$ was then forced to move on a $k$\,-dimensional surface ${\mathbb S}$ given by the algebraic 
equations $G_\alpha(q^A)=0$. 
Then equations of motion is known to follow from the modified Lagrangian, where the constraints are taken into account with help of auxiliary variables $\lambda_\alpha(t)$  as follows \cite{Arn_1, deriglazov2010classical}:  
\begin{eqnarray}\label{p1}
L(q^A, \dot q^A)-\lambda_\alpha G_\alpha(q^A). 
\end{eqnarray}
While we will discuss the case of an arbitrary surface ${\mathbb S}$, the most interesting  applications arise when the surface is a group manifold \cite{Arn_2, Mar_98, Hol_2007, Hol_1999, Dub_2001, Fom_2004}. Then in tangent space to the manifold ${\mathbb S}$ there is the natural basis composed by vector fields of the Lie algebra of the group, say, ${\bf X}_i$, with the Lie bracket $[{\bf X}_i, {\bf X}_j]=c_{ij}{}^k{\bf X}_k$. Then for any trajectory $q^A(t)\in{\mathbb S}$, we can write $\dot q^A=X^{A}{}_i\eta_i$. 

Let us outline and compare  three different possibilities to construct Hamiltonian formulation for the theory (\ref{p1}). 

{\bf (A)} The first possibility is to work with unconstrained variables. Let $x^i$ be local coordinates on ${\mathbb S}$, then (\ref{p1}) is known to be equivalent to the variational problem with the following unconstrained Lagrangian: $\tilde L(x^i, \dot x^i)\equiv L(q^A(x^i), \dot q^A(x^i))$. Denoting the conjugate momenta for $x^i$ by $p_i$, we immediately get the Hamiltonian equations $\dot x^i=\{x^i, H\}, \dot p_i=\{p_i, H\}$, with the canonical Poisson bracket $\{x^i, p_j\}=\delta^i_j$. In the case of the group manifold ${\mathbb S}$, this calculation at first glance completely ignores its group structure. 
However, Poincar\'e noticed long ago \cite{Poi_1901} that this is not the case. In the present context, his observation can be resumed as follows. In the phase space, one can pass to non-canonical variables, $(x^i, p_i)\rightarrow (x^i, \eta_i (x, p))$, in which the Poisson structure stores an information about the group structure of the manifold\footnote{The same observation turns out to be important in the semiclassical description of a spinning electron \cite{AAD_Rec, AAD_2019}  and of a photon \cite{AAD_2021_1}.}. Poincar\'e performed his calculations in terms of velocities, adjusting  infinitesimal variations of $\dot x^i$ and $\eta^i$, and obtained his equations in the Lagrangian-like form
\begin{eqnarray}\label{p2}
\frac{d}{dt}\frac{\partial\tilde L}{\partial\eta_i}=c_{ij}{}^k\frac{\partial\tilde L}{\partial\eta_j}\eta_k+f^i. 
\end{eqnarray}
Then Chetaev in \cite{Chet_1941} recognized the Hamiltonian character of these equations, pointing out the Hamiltonian and Poisson brackets.  In particular, for $SO(3)$ manifold, the resulting Hamiltonian equations are just the Euler equations of a rigid body \cite{Mar_98}.   

Note that the original problem  here is a mechanical system (\ref{p1}) with kinematic (that is velocity independent) constraints. So we expect that the Poincare-Chetaev equations should be obtained by direct application of the Dirac's procedure for constrained systems to this theory. This will be one of our tasks in the present work.  

We emphasize that the transition to independent variables is not always desirable. For instance, in the case of a rigid body, the $q^A$ variables are 9 elements of the matrix $R_{ij}$, subject to the orthogonality conditions. That is, to describe a rigid body, we need to know the evolution of $q^A$ and not $x^i$.
Besides, the description in terms of independent variables often turns out to be local, which can lead to misunderstandings, see \cite{AAD23_2}. Finally, solving equations for $q^A$ sometimes requires less effort than solving the same equations rewritten through $x^i$ \cite{Landau_8,AAD23_5}.

{\bf (B)} The second possibility is to work with the original variables using the Dirac's formalism \cite{Dir_1950, GT, deriglazov2010classical}.  We should pass to the Hamiltonian formulation introducing the conjugate momenta $p_A$ to all original variables $q^A$. The Hamiltonian equations then obtained with help of canonical Poisson bracket $\{q^A, p_B\}=\delta^A_B$ and of the Hamiltonian $H(q^A, p_B, \lambda_\alpha)$. Then the resulting equations depend on the auxilliary variables $\lambda_\alpha$.  The systematic method to exclude them is as follows. Besides the original constraints $G_\alpha=0$, in the Hamiltonian formulation arise certain highher-stage constraints $\Phi_\alpha=0$, and the complete set of constraints form the second-class system, that should be  taken into account  passing from the Poisson to Dirac bracket, say $\{q^A, p_B\}_D$. Writting equations of motion with help of Dirac bracket, it is known that the terms with $\lambda_\alpha$ in the Hamiltonian can simply be omitted. 

Once again, the underlying group structure in this formalism was at first sight ignored. 
To recognize it, there are two difficulties. First, the Dirac brackets are constructed for the excess number $n$ of momenta, as compared to the 
dimension $k$ of the Lie algebra. Therefore some reduction in the number of variables is needed. Second, to construct the Dirac bracket explicitly, it is necessary to invert the matrix composed of Poisson brackets of the constraints. For instance, in the case of $SO(3)$ group this is $12\times 12$\,-matrix. In the present work we slightly adjust the Dirac procedure (for the case of kinematic constraints), which alows us to avoid these two problems. To further clarify this issue, let us discuss the third known possibility to construct the Hamiltonian formulation. 

{\bf (C)}  In the case of a rigid body arises a kind of intermediate formalism between (A) and (B). Let us consider $SO(3)$ manifold with configuration-space variables being 9 elements of $3\times 3$\,-matrix subjected to six constraints $R^TR={\bf 1}$. Then the Hamiltonian equations are the Euler-Poisson equations, describing a free motion of an asymmetric rigid body \cite{Mac_1936, Lei_1965, Hol_2007, AAD23}
\begin{eqnarray}\label{p3} 
\dot R_{ij}=-\epsilon_{jkm}\Omega_k R_{im},   \qquad
I\dot{\boldsymbol\Omega}=[I{\boldsymbol\Omega}, {\boldsymbol\Omega}].   
\end{eqnarray}
Here $I$ is the inertia tensor, and the momenta $\Omega_i$ are the Hamiltonian counterpart of angular velocity in the body.  
There are 9 redundant coordinates $R_{ij}$, but only 3 independent momenta $\Omega_i$. So, if in case (A) we worked with independent set $(x^i, p_j)$, and in case (B) with redundant set $(q^A, p_B)$, then now we have an intermediate situation: $(q^A, p_j)$. 
As was shown in \cite{AAD23}, it is in this formalism that the Poincar\'e-Chetaev  equations can be obtained using the Dirac's method. However, 
in the calculations made in this work were used some specific properties of the group $SO(3)$. In Sect. 2  we show, how to construct the intermediate formalism in a more general case of an arbitrary surface, but for a special class of Lagrangians. Then in Sect. 3 this formalism will be used to obtain the Poincar\'e-Chetaev equations on $SO(3)$.

\section{Intermediate formalism for a special class of Lagrangians.}

In the  configuration space with the coordinates $q^A(t)$, $A=1, 2, \ldots n$, consider $k$\,-dimensional surface determined by functionally independent functions $G_\alpha(q^A)=0$. Without loss of generality, we assume that these equations can be resolved with respect to the first $n-k$\,-coordinates. In accordance to this, the set $q^A$ is divided on two subgroups, $q^\alpha$ and $q^i$. Greek indices from the beginning of the alphabet run from $1$ to $n-k$, while Latin indices from the middle of the alphabet run from $1$ to $k$. So 
\begin{eqnarray}\label{p3.1}
{\mathbb S}^k=\{q^A=(q^\alpha , q^i),  ~ G_\alpha(q^A)=0, ~ \det\left.\frac{\partial G_\alpha}{\partial q^\beta }\right|_{\mathbb S}=n-k, ~ \alpha=1, 2, ~ \ldots, n-k\}, 
\end{eqnarray}
and our variational problem is  (\ref{p1}). 
Applying the Dirac's method to the Lagrangian (\ref{p1}), we introduce conjugate momenta for all dynamical variables. Conjugate momenta 
for $\lambda_\alpha$ are the primary constraints: $p_{\lambda\alpha}=0$. Since the Lagrangian $L$ was assumed nonsingular, the expressions for $p_A$ can be resolved with respect to velocities: 
\begin{eqnarray}\label{p3.2}
p_A=\frac{\partial L}{\partial\dot q^A}\equiv \tilde f_A(q, \dot q), \qquad \mbox{then} \quad  \dot q^A=f^A(q, p), \quad \det\tilde f_{AB}\ne 0, 
\quad \mbox{where} \quad  \tilde f_{AB}\equiv\frac{\partial\tilde f_A}{\partial\dot q^B}.
\end{eqnarray} 
To find the Hamiltonian, we exclude the velocities from the expression $H=p_A\dot q^A-L+\lambda_\alpha G_\alpha+
v_\alpha p_{\lambda_\alpha}$, obtaining 
\begin{eqnarray}\label{p3.3}
H=p_A f^A(q, p)-L(q^A, f^B(q, p))+\lambda_\alpha G_\alpha(q^A)+v_\alpha p_{\lambda_\alpha}. 
\end{eqnarray}
By $v_\alpha $ we denoted the Lagrangian multipliers for the primary constraints.  
Preservation in time of the primary constraints, $\dot p_{\lambda_\alpha}=\{p_{\lambda_\alpha}, H\}=0$ implies $G_\alpha=0$ as the secondary constraints. In turn, the equation $dG_\alpha/dt=\{G_\alpha, H\}=0$ implies tertiary constraints, that should be satisfied by all true solutions 
\begin{eqnarray}\label{p6}
\Phi_{\alpha}\equiv G_{\alpha B}(q)f^B(q, p)=0, \qquad \mbox{where} \quad G_{\alpha B}\equiv\frac{\partial G_\alpha(q)}{\partial q^B}. 
\end{eqnarray}
The Lagrangian counterpart of this constraint is $\dot q^A\partial_A G_\alpha=0$, and mean that for true trajectories the velocity vector is 
tandent to the surface ${\mathbb S}$. 
Compute
\begin{eqnarray}\label{p7}
\mbox{rank}~\frac{\partial\Phi_{\alpha}}{\partial p_B}=\mbox{rank}~(G_{\alpha A}f^{AB})=n-k, \qquad \mbox{where} \quad  
f^{AB}=\frac{\partial f^A(q, p)}{\partial p_B}. 
\end{eqnarray}
This implies that the constraints $\Phi_\alpha$ are functionally independent and can be resolved with respect to some $n-k$ momenta of the set $p_A$. This implies also that the constraints  $G_\beta$ and $\Phi_\alpha$ are functionally independent. Computing their Poisson brackets we get the matrix
$\{ G_\alpha, \Phi_\beta \}=G_{\beta A}(q)f^{AB} G_{\alpha B}$.
Rank of this matrix can be analysed in the coordinates adapted with the surface: $q'^\alpha=G_\alpha(q^\alpha, q^i)$, $q'^i=q^i$. In this coordinates the surface is just the hyperplane $q'^\alpha=0$, then $G'_{\alpha A}=\delta_{\alpha A}$, and the matrix of brackets turn into $f'^{\beta\alpha}(q', p')$. It is $(n-k)\times (n-k)$ upper left block of the matrix $f^{AB}$, the latter is inverse of the Hessian matrix $\tilde f_{AB}$ of our theory.
We assume, that for our Lagrangian this matrix is nondegenerate
\begin{eqnarray}\label{p8}
\det f'^{\beta\alpha}(q', p')\ne 0.
\end{eqnarray}
This condition is satisfied, in particular, in the theories with a quadratic on velocities diagonalizable kinetic energy. Therefore, this may be not a very strong limitation for the applications.  

For such Lagrangians, our constraints $G_\beta$ and $\Phi_\alpha$ are of second class. 
Then preservation in time of the tertiary constraints gives fourth-stage constraints that involve $\lambda_\alpha$, and can be used to find them through $q^A$ and $p_A$. At last, preservation in time of the fourth-stage constraints gives an equation that determines the Lagrangian 
multipliers $v_\alpha$. We do not write out these equations, we will not need them.

To proceed further, let us construct noncanonical phase-space coordinates  with special properties. The matrix $G_{\alpha B}$ of Eq. (\ref{p6}) is composed by $(n-k)$ linearly independent vector fields ${\bf G}_\alpha$, orthogonal to the surface ${\mathbb S}$. The linear 
system $G_{\alpha B} x_B=0$ has a general solution\footnote{To avoid a posible confusion, we point out that in the similar Eq. (\ref{p6}), representing the tertiary constraints, $f^A$ are given functions of $q$ and $p$.} of the form $x_B=c^i G_{i B}$, where the linearly independent vectors ${\bf G}_i$ are fundamental solutions to this system
\begin{eqnarray}\label{p9}
{\bf G}_i=( ~G_{i1}(q), G_{i2}(q), \ldots , G_{i, n-k}(q), 0, \ldots , 1, 0 \ldots , 0 ~ ), \qquad G_{\alpha B}G_{i B}=0. 
\end{eqnarray}
By construction, these vector fields form a basis of tangent space to the surface ${\mathbb S}$. Together with ${\bf G}_\alpha$, they form a basis of tangent space to the entire configuration space. Using the rows ${\bf G}_\beta$ and ${\bf G}_j$, we construct an invertible matrix $G_{BA}$, and use it to define the new momenta $\pi_B$
\begin{eqnarray}\label{p10}
G_{BA}(q)=\left(
\begin{array}{c}
G_{\beta A} \\
G_{j A}
\end{array}
\right), \qquad \pi_B=G_{BA}(q) p_A, \quad \mbox{then} \quad p_A=G^{-1}_{AB}(q)\pi_B\equiv\tilde G_{AB}(q)\pi_B. 
\end{eqnarray}
Let us take $q^A$ and $\pi_A$ as the new phase-space coordinates. 
Their special property is that both $q^A$ and $\pi_i$ have vanishing brackets with the original constraints $G_\alpha$
\begin{eqnarray}\label{p11}
\{ q^A, G_\alpha \}=0, \qquad \{\pi_i, G_\alpha \}=0, 
\end{eqnarray}
the latter equality is due to Eq. (\ref{p9}). 

Let us rewrite our theory in the new variables.  Using the canonical brackets $\{q^A, p_B\}=\delta^A{}_B$, we get Poisson brackets of the new variables
\begin{eqnarray}\label{p12}
\{ q^A, q^B\}=0, \qquad \{q^A, \pi_B \}=G_{BA}(q), \qquad \{\pi_A, \pi_B \}=-c_{AB}{}^D(q)\tilde G_{DE}(q)\pi_E, 
\end{eqnarray}
where appeared the Lie brackets of basic vector fields ${\bf G}_A$
\begin{eqnarray}
c_{AB}{}^D=[{\bf G}_A, {\bf G}_B]^D=G_{AE}\partial_E G_{BD}-G_{BE}\partial_E G_{AD}, \label{p13}\\
c_{ij}{}^k=0. \qquad \qquad \qquad \qquad \qquad \qquad \label{p13.1}
\end{eqnarray} 
Therefore the Lie bracket of the vector fields ${\bf G}_A$ determines Poisson structure of our theory in the sector $\pi_A$. 
The structure functions $c_{ij}{}^k$ vanish for our choice of basic vectors ${\bf G}_i$ of special form, see Eq. (\ref{p9}).  The Hamiltonian (\ref{p3.3}) reads
\begin{eqnarray}\label{p14}
H=\tilde G_{AC}\pi_C f^A(q, \tilde G\pi)-L(q^A, f^B(q, \tilde G\pi))+\lambda_\alpha G_\alpha(q^A). 
\end{eqnarray}
At last, our second-class constraints in the new coordinates are
\begin{eqnarray}\label{p15}
G_\alpha (q^A)=0, \qquad \Phi_\alpha\equiv G_{\alpha A}(q)f^A(q, \tilde G\pi)=0.
\end{eqnarray}
Using them, we construct the Dirac bracket
\begin{eqnarray}\label{p16}
\{ A, B\}_D=\{A, B\}-\{A, T^i\}\triangle^{-1}_{ij}\{T^j, B\}.
\end{eqnarray}
Here $T^i$ is the set of all constraints: $T^i=(G_\alpha, \Phi_\beta)$. Besides, denoting symbolically the blocks $b=\{ G, \Phi \}$ 
and $c=\{ \Phi, \Phi  \}$, the matrices $\triangle$ and $\triangle^{-1}$ are   
\begin{eqnarray}\label{p17}
\triangle=\left(
\begin{array}{cc}
0 & b \\
-b^T& c 
\end{array}\right), \qquad 
\triangle^{-1}=\left(
\begin{array}{cc}
b^{-1 T}cb^{-1} & -b^{-1 T} \\
b^{-1} & 0 
\end{array}\right).
\end{eqnarray}
This implies the following structure of the Dirac bracket
\begin{eqnarray}\label{p18}
\{ A, B\}_D=\{A, B\}-\{A, G\}\triangle ' \{G, B\}+\{A, G\}\triangle '' \{\Phi, B\}.
\end{eqnarray}
Taking into account Eqs. (\ref{p11}), we conclude that in the passage from Poisson bracket (\ref{p12}) to the Dirac bracket, the brackets (\ref{p12}) of the basic variables $q^A$ and $\pi_i$ will not be modified, retaining their original form. So, fortunately, we do not need to calculate the explicit form of the matrix $\triangle^{-1}$ appeared in (\ref{p16}). The constraint's functions (\ref{p15}) are Casimir functions of the Dirac bracket (\ref{p16}). 

Let us confirm that the tertiary constraints $\Phi_\alpha$ from (\ref{p15}) can be resolved with respect to $\pi_\alpha$. To this aim we compute $\det(\partial\Phi_\alpha/\partial\pi_\beta)$ in the adapted coordinates, and show that it is not zero
\begin{eqnarray}\label{p19}
\det\frac{\partial(G'_{\alpha A}f'^A(q', \tilde G'\pi')}{\partial\pi'_\beta}=\det [G'_{\alpha A}f'^{AD}(q', \tilde G'\pi')\tilde G'_{D \beta}]=\det f'^{\alpha\beta}(q', \pi' )\ne 0.
\end{eqnarray}
Here we used that in adapted coordinates $G'_{\alpha A}=(\delta_{\alpha\beta}, {\bf 0})$ and $\tilde G'_{D \beta}=(\delta_{\alpha\beta}, {\bf 0})^T$. It is not zero for our class of Lagrangians (\ref{p8}). 

The formulation of the theory in terms of Dirac bracket makes it much more transparent. Indeed, according to the Dirac's formalism, we now can omite all terms with constraints in the Hamiltonian. Besides, we can use the constraints before the calculation of the brackets. Therefore, resolving the constraints (\ref{p15}) for $\pi_\alpha$ and excluding them from the formalism, we get the desired intermediate formulation of our theory in terms of $q^A$ and $\pi_i$.

In particular, from Eqs. (\ref{p12}) we get the Poisson structure of the intermediate formulation as follows: 
\begin{eqnarray}\label{p20}
\{ q^A, q^B \}_D=0, \qquad \{q^\alpha, \pi_i \}_D=G_{i \alpha}(q), \qquad \{q^j, \pi_i \}_D=\delta^j{}_i,  \qquad 
\{\pi_i, \pi_j \}_D=-c_{ij}{}^\alpha[\tilde G_{\alpha k}\pi_k+\tilde G_{\alpha\beta}\pi_\beta(q^A, \pi_i)], 
\end{eqnarray}
where $\pi_\beta(q^A, \pi_i)$ is the solution to the tertiary constraints $\Phi_\alpha=0$. In general, the brackets are nonlinear for both $q^A$ and $\pi_i$. Their dependence on the choice of tangent vector fields ${\bf G}_i$ to the surface ${\mathbb S}$ is encoded in three places: in the 
brackets $\{q^\alpha, \pi_j\}$, in the matrix $\tilde G$, as well as in the structure functions $c_{ij}{}^\alpha$, see Eq. (\ref{p13}). 

In the Hamiltonian (\ref{p14}), we omite the term containing the constraints $G_\alpha$, and exclude $\pi_\alpha$. Let us denote the resulting expression by $H_0(q^A, \pi_j)$.  
Hamiltonian equations of intermediate formalism are obtained with use of Dirac brackets: $\dot q^A=\{q^A, H_0( q^B, \pi_j) \}_D$, $\dot\pi_i=\{\pi_i, H_0( q^B, \pi_j) \}_D$. 

Being one of the classical problems in the theory of integrable systems and classical mechanics, these issues could be of interest in the modern studies of various aspects related with construction and behavior of spinning particles and rotating bodies in external fields beyond the pole--dipole approximation \cite{Joon_21.1,Joon_23.2,Far_23.3,Joo_23.4,Dmi_2023,Tao_2023,Yas_2023}.

\section{Poincar\'e-Chetaev equations on $SO(3)$.}      

Let us see, how the intermediate formalism of previous section works in the case of $SO(3)$\,-manifold. The detailed computations of $SO(3)$\,-case were presented in \cite{AAD23}, so here we only outline the relationship between these calculations and the intermediate formalism. 

First we note that the matrix $G_{BA}$ can equally be used  to construct another coordinates 
\begin{eqnarray}\label{p21.0}
\pi_\alpha=G_{\alpha B}f^B\equiv \Phi_\alpha, \qquad \pi_i=G_{i B}p_B, 
\end{eqnarray}  
which contain the constraints $\Phi_\alpha$ as a part of new momenta. This change of variables is equivalent to that used in previous section, but turns out to be  more convenient in the case of $SO(3)$. 

The rotational degrees of 
freedom $R_{ij}(t)$ of a rigid body are determined by the Lagrangian action
\begin{eqnarray}\label{p21}
S=\int dt ~ ~ \frac12 g_{ij}\dot R_{ki}\dot R_{kj} -\frac12 \lambda_{ij}\left[R_{ki}R_{kj}-\delta_{ij}\right].
\end{eqnarray} 
with the universal initial conditions $R_{ij}(0)=\delta_{ij}$. Their conjugate momenta are denoted by $p_{ij}=\partial L/\partial\dot R_{ij}$. In the expression (\ref{p21}),  the mass matrix is taken to be diagonal: $g_{ij}= diagonal ~ (g_1, g_2, g_3)$, and is related with the inertia tensor as follows:  $2g_1=I_2+I_3-I_1$, $2g_2=I_1+I_3-I_2$, $2g_3=I_1+I_2-I_3$. The Hamiltonian of the theory reads
\begin{eqnarray}\label{p22}
H=\frac12 g^{-1}_{ij}p_{ki}p_{kj}+\frac12 \lambda_{ij}[R_{ki}R_{kj}-\delta_{ij}].  
\end{eqnarray}
Applying the formalism of previous section and comparing it with the calculations of Sect. XII of the work \cite{AAD23}, we have the following table for identification of the basic quantities: 
\begin{eqnarray}\label{p23}
q^A \sim R_{ij}, \qquad p_A \sim p_{ij}, \qquad \tilde f^A \sim \dot R_{ik}g_{ki}, \qquad f_A \sim p_{ik}g^{-1}_{kj}, \label{p22} \\
\pi_\alpha=\Phi_{\alpha} \sim {\mathbb P}_{(ij)}=\frac12[R^Tpg^{-1}+(R^Tpg^{-1})^T]_{ij},  \label{p22.1} \\
\pi_i \sim \hat\Omega_{ij}=-\frac12[R^Tpg^{-1}-(R^Tpg^{-1})^T]_{ij} \sim M_n=-I_{nk}\epsilon_{kij}(R^Tp)_{ij}. \label{p22.1}
\end{eqnarray}
In the case of $SO(3)$,  the final results acquire more simple form, if we use the angular momentum in the body $M_n=(I\Omega)_n$ instead of the angular velocity in the 
body $\hat\Omega_{ij}$. Therefore the identification of new coordinates of general scheme with $SO(3)$\,-case 
is as follows: $(q^A, \pi_\alpha, \pi_i) \sim (R_{ij}, {\mathbb P}_{(ij)}, M_i)$. Making calculations of the previous section in these coordinates, we get the following final Hamiltonian and the brackets 
\begin{eqnarray}\label{p24}
H_0=\frac12 I^{-1}_{ij}M_i M_j;   
\end{eqnarray}
\begin{eqnarray}\label{11.20}
\{ R_{ij}, R_{ab}\}=0, \qquad \{M_i, M_j\}=-\epsilon_{ijk}((R^TR)^{-1}{\bf M})_k, \qquad 
\{M_i, R_{jk}\}=-\epsilon_{ikm}R^{-1 T}_{jm};  
\end{eqnarray}
Using the orthogonality constraint on r.h.s. of the brackets (\ref{11.20}), we obtain more simple expressions
\begin{eqnarray}\label{11.20.2}
\{ R_{ij}, R_{ab}\}=0, \qquad \{M_i, M_j\}=-\epsilon_{ijk}M_k, \qquad 
\{M_i, R_{jk}\}=-\epsilon_{ikm}R_{jm}.  
\end{eqnarray}
By direct computations, it can be verified that they still satisfy the Jacobi identity and lead to the same equations (\ref{p26}).  They  were suggested by Chetaev \cite{Chet_1941} as the posible Poisson structure corresponding to the Euler-Poisson equations.

Denoting the rows of the matrix $R_{ij}$ by ${\bf a}$, ${\bf b}$ and ${\bf c}$, the bracket (\ref{11.20.2}) reads: $\{M_i, a_j\}_D=-\epsilon_{ijk}a_k$,  $\{M_i, b_j\}_D=-\epsilon_{ijk}b_k$, $\{M_i, c_j\}_D=-\epsilon_{ijk}c_k$. Therefore the Poisson structure of a rigid body can be identified with semidirect sum of the algebra $\bar{so}(3)$ with three translation algebras. The brackets (\ref{11.20.2}) were suggested by Chetaev \cite{Chet_1941} as the posible Poisson structure corresponding to the Euler-Poisson equations.

Using  the rule: $\dot z=\{ z, H_0 \}_D$, we get the Euler-Poisson equations 
\begin{eqnarray}\label{p26} 
\dot R_{ij}=-\epsilon_{jkm}(I^{-1}M)_k R_{im}, 
\end{eqnarray}
\begin{eqnarray}\label{p27}
\dot{\bf M}=[{\bf M}, I^{-1}{\bf M}].   
\end{eqnarray}
Here the bracket $[~, ~]$ means the vector product of ${\mathbb R}^3$. 
Using the rows ${\bf a}$, ${\bf b}$ and ${\bf c}$, Eqs. (\ref{p26}) can be separated: $\dot{\bf a}=[{\bf a}, I^{-1}{\bf M}]$, 
$\dot{\bf b}=[{\bf b}, I^{-1}{\bf M}]$  and $\dot{\bf c}=[{\bf c}, I^{-1}{\bf M}]$. Then the entire system  (\ref{p26}), (\ref{p27}) breaks down into three. For instance, in the case of the row ${\bf a}$ we have
\begin{eqnarray}\label{p28} 
\dot a_i=[{\bf a}, I^{-1}{\bf M}]_i, \qquad \dot M_j=[{\bf M}, I^{-1}{\bf M}]_j,  
\end{eqnarray}
$\{M_i, a_j\}_D=-\epsilon_{ijk}a_k$ in accordance with (\ref{11.20.2}) and. For this unconstrained Hamiltonian system we can use the known formula of Hamiltonian mechanics to write solutions to the equations (\ref{p28}) in terms of exponential of the Hamiltonian vector field \cite{AAD_2022}. Doing the same for the vectors ${\bf b}$ and ${\bf c}$, we get the solution to Euler-Poisson equations (\ref{p26}) and (\ref{p27}) as follows: 
\begin{eqnarray}\label{p29}
M_i(t, M_{0k})=e^{t[{\bf M}_0, I^{-1}{\bf M}_0]_j\frac{\partial}{\partial M_{0j}}}M_{0i}, \quad 
a_i(t, M_{0k})=e^{t([{\bf M}_0, I^{-1}{\bf M}_0]_j\frac{\partial}{\partial M_{0j}}+[{\bf a}_0, I^{-1}{\bf M}_0]_j\frac{\partial}{\partial a_{0j}})}a_{0i}, \cr 
b_i(t, M_{0k})=e^{t([{\bf M}_0, I^{-1}{\bf M}_0]_j\frac{\partial}{\partial M_{0j}}+{\bf b}_0, I^{-1}{\bf M}_0]_j\frac{\partial}{\partial b_{0j}})}b_{0i}, \quad 
c_i(t, M_{0k})=e^{t([{\bf M}_0, I^{-1}{\bf M}_0]_j\frac{\partial}{\partial M_{0j}}+[{\bf c}_0, I^{-1}{\bf M}_0]_j\frac{\partial}{\partial c_{0j}})}c_{0i}. \quad 
\end{eqnarray}
It depends on three arbitrary constants $M_{0k}$,  and is therefore a general solution to the Euler-Poisson equations (recall that for the rotation matrix the initial conditions are universal: $R_{ij}(0)=\delta _{ij}$).

\begin{acknowledgments}
The work has been supported by the Brazilian foundation CNPq (Conselho Nacional de Desenvolvimento Cient\'ifico e Tecnol\'ogico - Brasil). 
\end{acknowledgments}

\end{document}